\documentclass[a4paper,10pt]{article}

\usepackage{epsf}
\usepackage{graphicx}    % Include figure files

\newcommand{\be}[1]{\begin{equation}\label{#1}}
\newcommand{\ee}{\end{equation}}
\newcommand{\bea}{\begin{eqnarray}}
\newcommand{\eea}{\end{eqnarray}}

\def\gsim{ \lower .75ex \hbox{$\sim$} \llap{\raise .27ex \hbox{$>$}} }
\def\lsim{ \lower .75ex \hbox{$\sim$} \llap{\raise .27ex \hbox{$<$}} }

\renewcommand{\markright}{\markright{\thepage}}

\begin{document}

\begin{titlepage}

%\begin{flushright}
%astro-ph/0609597
%\end{flushright}

\vspace{5mm}

\begin{center}

{\Large \bf Statefinder diagnosis for the interacting model of
holographic dark energy}

\vspace{10mm}

{\large Jingfei Zhang,$^{1}$ Xin Zhang,$^{2}$ and Hongya Liu$^{1}$}

\vspace{5mm} {\em $^{1}$School of Physics and Optoelectronic
Technology, Dalian University of Technology, Dalian 116024, People's
Republic of China\\
$^{2}$Kavli Institute for Theoretical Physics China, Institute of
Theoretical Physics, Chinese Academy of Sciences (KITPC/ITP-CAS),
P.O.Box 2735, Beijing 100080, People's Republic of China}

\end{center}

\vspace{5mm}
\begin{abstract}
In this paper, we investigate the holographic dark energy model with
interaction between dark energy and dark matter, from the
statefinder viewpoint. We plot the trajectories of the interacting
holographic dark energy model for different interaction cases as
well as for different values of the parameter $c$ in the
statefinder-plane. The statefinder diagrams characterize the
properties of the holographic dark energy and show the
discrimination between the two cases with and without interaction.
As a result, we show the influence of the interaction on the
evolution of the universe in the statefinder diagrams. Moreover, as
a complement to the statefinder diagnosis, we study the interacting
holographic dark energy model in the $w-w'$ plane, which can provide
us with a dynamical diagnosis.

%\noindent PACS numbers: 95.36.+x, 98.80.Es, 98.80.-k
\end{abstract}

\end{titlepage}

\newpage

\setcounter{page}{2}

Today it has been confirmed that our universe is undergoing an
accelerating expansion through numerous cosmological observations,
such as type Ia supernovae (SNIa) \cite{SN}, large scale structure
(LSS) \cite{LSS} and cosmic microwave background (CMB) \cite{CMB}.
This cosmic acceleration is attributed to a mysterious dominant
component, dark energy, with negative pressure. The combined
analysis of cosmological observations suggests that the universe is
spatially flat, and consists of about $70\%$ dark energy, $30\%$
dust matter, and negligible radiation. Many candidates have been
proposed to interpret or describe the properties of dark energy,
though its nature still remains enigmatic. The most obvious
theoretical candidate of dark energy is the cosmological constant
$\lambda$ \cite{Einstein:1917,cc} which has the equation of state
$w=-1$. However, as is well known, there are two difficulties arise
from the cosmological constant scenario, namely the two famous
cosmological constant problems --- the ``fine-tuning'' problem and
the ``cosmic coincidence'' problem \cite{coincidence}. Theorists
have made lots of efforts to try to resolve the cosmological
constant problem but all these efforts were turned out to be
unsuccessful.

Also, there is an alternative proposal to dark energy --- the
dynamical dark energy scenario. The dynamical dark energy scenario
is often realized by some scalar field mechanism which suggests that
the energy form with negative pressure is provided by a scalar field
evolving down a proper potential. A lot of scalar-field dark energy
models have been studied, including quintessence
\cite{quintessence}, K-essence \cite{kessence}, tachyon
\cite{tachyon}, phantom \cite{phantom}, ghost condensate
\cite{ghost} and quintom \cite{quintom} etc.. In addition, other
proposals on dark energy include scenarios of interacting dark
energy \cite{intde}, braneworld \cite{brane}, Chaplygin gas
\cite{cg}, and so forth. By far, obviously, it is not yet clear if
dark energy is a cosmological constant or a dynamical field.
Generally, theorists believe that we can not entirely understand the
nature of dark energy before a complete theory of quantum gravity is
established \cite{Witten:2000zk}.

However, in this circumstance, we still can make some efforts to
probe the properties of dark energy according to some principle of
quantum gravity. The holographic dark energy model is an example of
such effort, which stems from the holographic principle and can
provide us with an intriguing way to interpret the dynamics of dark
energy. The holographic principle is an important result of the
recent researches of exploring the quantum gravity and is
enlightened by investigations of the quantum property of black holes
\cite{holoprin}. According to the holographic principle, the number
of degrees of freedom for a system within a finite region should be
finite and should be bounded roughly by the area of its boundary. In
the cosmological context, the holographic principle will set an
upper bound on the entropy of the universe. Motivated by the
Bekenstein entropy bound, it seems plausible that one may require
that for an effective quantum field theory in a box of size $L$ with
UV cutoff $\Lambda$, the total entropy should satisfy
$S=L^3\Lambda^3\leq S_{BH}\equiv\pi M_{\rm P}^2L^2$, where $S_{BH}$
is the entropy of a black hole with the same size $L$. However,
Cohen et al. \cite{Cohen:1998zx} pointed out that to saturate this
inequality some states with Schwartzschild radius much larger than
the box size have to be counted in. As a result, a more restrictive
bound, the energy bound, has been proposed to constrain the degrees
of freedom of the system, requiring the total energy of a system
with size $L$ not to exceed the mass of a black hole with the same
size, namely, $L^3\Lambda^4=L^3\rho_{\rm de}\leq L M_{\rm P}^2$.
This means that the maximum entropy is in the order of
$S_{BH}^{3/4}$. When we take the whole universe into account, the
vacuum energy related to this holographic principle is viewed as
dark energy, usually dubbed holographic dark energy. The largest IR
cut-off $L$ is chosen by saturating the inequality, so that we get
the holographic dark energy density
\begin{equation}
\rho_{\rm de}=3c^2M_{\rm P}^2L^{-2}~,\label{de}
\end{equation}
where $c$ is a numerical constant\footnote{The parameter $c$ is
introduced to parameterize some uncertainties, such as the species
of quantum fields in the universe, the effect of curved spacetime,
and so forth.} (note that $c>0$ is assumed), and as usual $M_{\rm
P}$ is the reduced Planck mass. If we take $L$ as the size of the
current universe, for instance the Hubble scale $H^{-1}$, then the
dark energy density will be close to the observed value. However,
Hsu \cite{Hsu:2004ri} pointed out that this yields a wrong equation
of state for dark energy. Li \cite{Li:2004rb} subsequently proposed
that the IR cutoff $L$ should be given by the future event horizon
of the universe,
\begin{equation}
R_{\rm eh}(a)=a\int\limits_t^\infty{dt'\over
a(t')}=a\int\limits_a^\infty{da'\over Ha'^2}~.\label{eh}
\end{equation}
Such a holographic dark energy looks reasonable, since it may
provide simultaneously natural solutions to both dark energy
problems, as demonstrated in Ref. \cite{Li:2004rb}. Meanwhile, other
applications of the holographic principle in cosmology
\cite{otherholo} show that holography is an effective way to
investigate cosmology. For other extensive studies, see e.g.
\cite{holoext}--\cite{Wang:2005jx}.

Besides, some interacting models are discussed in many works because
these models can help to understand or alleviate the coincidence
problem by considering the possible interaction between dark energy
and cold dark matter due to the unknown nature of dark energy and
dark matter. In addition, the proposal of interacting dark energy is
compatible with the current observations such as the SNIa and CMB
data \cite{Guo:2007zk}. For the interacting model of holographic
dark energy see \cite{Wang:2005jx}.

On the other hand, since more and more dark energy models have been
constructed for interpreting or describing the cosmic acceleration,
the problem of discriminating between the various contenders is
becoming emergent. In order to be capable of differentiating between
those competing cosmological scenarios involving dark energy, a
sensitive and robust diagnosis for dark energy models is a must. In
addition, for some geometrical models arising from modifications to
the gravitational sector of the theory, the equation of state no
longer plays the role of a fundamental physical quantity, so it
would be very useful if we could supplement it with a diagnosis
which could unambiguously probe the properties of all classes of
dark energy models. For this purpose a diagnostic proposal that
makes use of parameter pair $\{r,s\}$, the so-called ``statefinder",
was introduced by Sahni et al. \cite{sahni}. The statefinder probes
the expansion dynamics of the universe through higher derivatives of
the scale factor $\stackrel{...}{a}$ and is a ``geometrical''
diagnosis in the sense that it depends on the scale factor and hence
on the metric describing space-time. Since different cosmological
models involving dark energy exhibit different evolution
trajectories in the $s-r$ plane, the statefinder can be used to
diagnose different dark energy models \cite{Alam:2003sc}.

In this paper, we focus on a model of holographic dark energy with
interaction between dark energy and dark matter and study the
influence of the interaction to the cosmic evolution. Moreover, we
use the statefinder to diagnose various cases with different
interaction strength and different parameter $c$ in the holographic
model.

%\section{Holographic dark energy}
Let us start with a spatially flat Friedmann-Robertson-Walker (FRW)
universe with dust matter and holographic dark energy. The Friedmann
equation reads
\begin{equation}
3M_{\rm P}^{2}H^{2}\label{f1}=\rho_{\rm de}+\rho_{\rm m},
\end{equation}
where $\rho_{\rm m}$ is the energy density of matter and $\rho_{\rm
de}=3c^2M_{\rm P} ^2R_{\rm h}^{-2}$ is the dark energy density. The
total energy density satisfies a conservation law,
\begin{equation}
\dot{\rho}_{\rm de}+\dot{\rho}_{\rm m}=-3H(\rho+P),\label{f3}%
\end{equation}
where $\rho=\rho_{\rm m}+\rho_{\rm de}$ is the total energy density
of the universe, and $P=P_{\rm de}=w\rho_{\rm de}$ is the total
pressure ($w$ denotes the equation of state of dark energy). Note
that since the matter component is mainly contributed by the cold
dark matter, we ignore the contribution of the baryon matter here
for simplicity. By introducing $\Omega_{\rm de}=\rho_{\rm
de}/(3M_{\rm P} ^{2}H^{2})$ and $\Omega_{\rm m}=\rho_{\rm
m}/(3M_{\rm P}^{2}H^{2})$, the Friedmann equation can also be
written as $\Omega_{\rm de}+\Omega_{\rm m}=1$. Furthermore, if we
proceed to consider a scenario of interacting dark energy,
$\rho_{\rm m}$ and $\rho_{\rm de}$ do not satisfy independent
conservation laws, they instead satisfy
\begin{equation}
\dot{\rho}_{\rm m}+3H\rho_{\rm m}=Q, \label{a1}
\end{equation}
and
\begin{equation}
\dot{\rho}_{\rm de}+3H(1+w)\rho_{\rm de}=-Q,\label{a2}
\end{equation}
where $Q$ describes the interaction between dark energy and dark
matter. It is obvious that the interaction term $Q$ could not be
introduced by considering some micro-process currently, so a
phenomenological way is the must. One possible choice for the
interaction term is setting
\begin{equation}
Q=3b^2 H\rho, \label{Q}
\end{equation}
where $b$ is a constant describing the coupling strength. This
expression for the interaction term was first introduced in the
study of the suitable coupling between a quintessence scalar field
and a pressureless cold dark matter component, in order to get a
scaling solution to the coincidence problem \cite{Pavon:2005yx}.

Taking the ratio of energy densities as $\mu=\rho_{\rm m}/\rho_{\rm
de}$ and using the Friedmann equation $\Omega_{\rm de}+\Omega_{\rm
m}=1$, we have $\mu=(1-\Omega_{\rm de})/\Omega_{\rm de}$ and
$\dot{\mu}=-\dot{\Omega}_{\rm de}/\Omega_{\rm de}^2$. Furthermore,
from (\ref{a1}), (\ref{a2}) and (\ref{Q}), we obtain
\begin{equation}
 \dot{\mu}=3b^2
H(1+\mu)^2+3H\mu w.\label{3}
\end{equation}
Combining these results, we easily get the equation of state of dark energy
\begin{eqnarray}
w & = &\frac{-\dot{\Omega}_{\rm de}/\Omega_{\rm de}^2-3b^2H(1+\mu)^2}{3H\mu}  \nonumber \\
& = &-\frac{\Omega_{\rm de}'}{3\Omega_{de}(1-\Omega_{\rm
de})}-\frac{b^2}{\Omega_{\rm de}(1-\Omega_{\rm de})}, \label{4}
\end{eqnarray}
where prime denotes the derivative with respect to $x=\ln a$.

Using the definition of the holographic dark energy (\ref{de}) and
the Friedmann equation, the future event horizon (\ref{eh}) can be
expressed as $R_{\rm h}=c\sqrt{1+\mu}/H$. Then, for this expression,
taking the derivative with respect to $t$ and reducing the result,
we get
\begin{equation}
\frac{\Omega_{\rm de}'}{\Omega_{\rm de}^2}=(1-\Omega_{\rm
de})\left[\frac{1}{\Omega_{\rm de}}+ \frac{2}{c\sqrt{\Omega_{\rm
de}}}-\frac{3b^2}{\Omega_{\rm de}(1-\Omega_{\rm de})}\right].
\label{5}
\end{equation}
It is notable that this differential equation governs the whole
dynamics of the interacting model of holographic dark energy.
Substituting (\ref{5}) to (\ref{4}) yields
\begin{equation}
w= -\frac{1}{3}-\frac{2\sqrt{\Omega_{\rm
de}}}{3c}-\frac{b^2}{\Omega_{\rm de}} .\label{w}
\end{equation}
Then we can compute the deceleration parameter
\begin{eqnarray}
q& = & -\frac{\ddot{a}}{aH^2}=\frac{1}{2}+\frac{3}{2}w\Omega_{\rm
de} =\frac{1}{2}\left(1-3b^2-\Omega_{\rm de}-\frac{2}{c}\Omega_{\rm
de}^{\frac{3}{2}}\right).\label{6}
\end{eqnarray}

\begin{figure}[hbtp]
\begin{center}
\includegraphics[width=10cm]{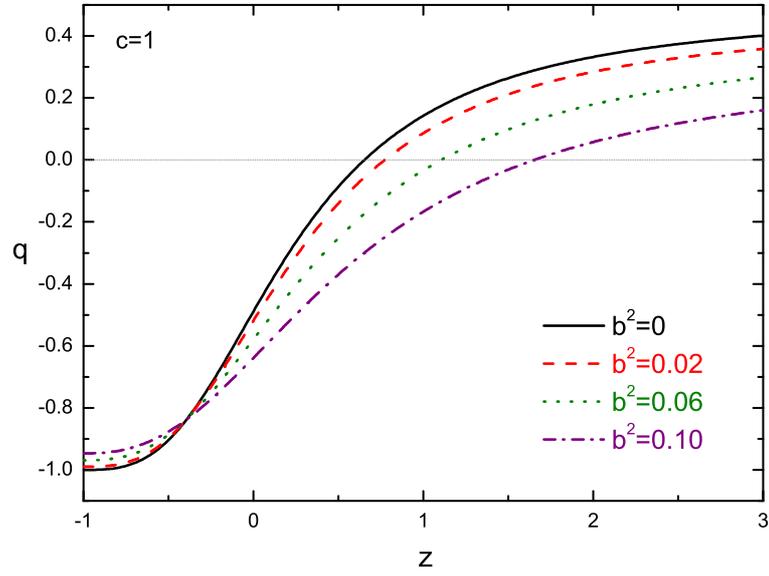}
\end{center}
\caption{Evolution of the deceleration parameter $q$ with a fixed
parameter $c$. In this plot, we take $c=1$, $\Omega_{\rm de0}=0.73$,
and vary $b^2$ as 0, 0.02, 0.06, and 0.10, respectively.}
\end{figure}

\begin{figure}[hbtp]
\begin{center}
\includegraphics[width=10cm]{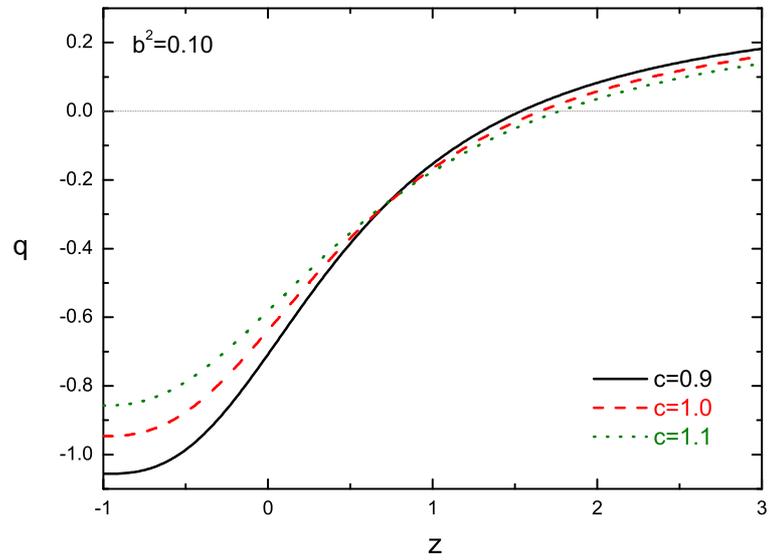}
\end{center}
\caption{Evolution of the deceleration parameter $q$ with a fixed
coupling $b^2$. In this plot, we take $b^2=0.10$, $\Omega_{\rm
de0}=0.73$, and vary $c$ as 0.9, 1.0 and 1.1, respectively.}
\end{figure}

In order to show the influence of interaction to the cosmic
evolution, the cases with dependence of the parameter $b^2$ for the
deceleration parameter $q$ are shown in Fig. 1. In Fig. 1, we fix
$c=1$ and take the coupling constant $b^2$ as 0, 0.02, 0.06, and
0.10, respectively. Besides, the cases with a fixed $b^2$ and
various values of $c$ are also interesting. In Fig. 2, fixing the
coupling constant $b^2=0.10$, we plot the evolution diagram of the
deceleration parameter $q$ with different values of parameter $c$
(here we take the values of $c$ as 0.9, 1.0, and 1.1, respectively).
From Figs. 1 and 2 we learn that the universe experienced an early
deceleration and a late time acceleration. Fig. 1 shows that, for a
fixed parameter $c$, the cosmic acceleration starts earlier for the
cases with interaction than the ones without coupling (for this
point see also, e.g., \cite{Amendola:2002kd}). Moreover, the
stronger the coupling between dark energy and dark matter is the
earlier the acceleration of universe began. However, the cases with
smaller coupling will get bigger acceleration finally in the far
future. In addition, Fig. 2 shows that the acceleration starts
earlier when $c$ is larger for the same coupling $b^2$, but finally
a smaller $c$ will lead to a bigger acceleration. It should be
pointed out that, in the interacting holographic dark energy model,
the interaction strength has an upper limit because of the
evolutionary behavior of the holographic dark energy. For detailed
discussions about correlation of the coupling $b^2$ and the
parameter $c$, see \cite{Wang:2005jx}. It is remarkable that, with
the interaction between dark energy and dark matter, the case of
$c=1$ could not enter a de Sitter phase in the infinite future. In
short, the influence of the interaction between dark energy and dark
matter to the cosmic evolution is obvious, as manifested by Figs. 1
and 2. On the other hand, nevertheless, as Eq. (\ref{6}) shows,
though the deceleration parameter $q$ carries the information of the
equation of state of dark energy $w$, the property of dynamical
evolution for $w$ can not be read out from $q$. For diagnosing
properties and evolutionary behaviors of dark energy models
exquisitely, more powerful diagnostic tool is a must.

%\section{The interacting model of holographic dark energy}

%\section{Statefinder diagnostic}
Now we turn to the statefinder diagnosis. For characterizing the
expansion history of the universe, one defines the geometric
parameters $H=\dot{a}/a$ and $q=-\ddot{a}/aH^2$, namely the Hubble
parameter and the deceleration parameter. It is clear that
$\dot{a}>0$ means the universe is undergoing an expansion and
$\ddot{a}>0$ means the universe is experiencing an accelerated
expansion. From the cosmic acceleration, $q<0$, one infers that
there may exist dark energy with negative equation of state,
$w<-1/3$ and likely $w\sim -1$, but it is hard to deduce the
information of the dynamical property of $w$ (namely the time
evolution of $w$) from the value of $q$. In order to extract the
information on the dynamical evolution of $w$, it seems that we need
the higher time derivative of the scale factor, ${\stackrel{...}a}$.
Another motivation for proposing the statefinder parameters stems
from the merit that they can provide us with a diagnosis which could
unambiguously probe the properties of all classes of dark energy
models including the cosmological models without dark energy
describing the cosmic acceleration. Though at present we can not
extract sufficiently accurate information of $\ddot{a}$ and
${\stackrel{...}a}$ from the observational data, we can expect,
however, the high-precision observations of next decade may be
capable of doing this. Since different cosmological models exhibit
different evolution trajectories in the $s-r$ plane, the statefinder
parameters can thus be used to diagnose the evolutionary behaviors
of various dark energy models and discriminate them from each other.
In this paper, we apply the statefinder diagnosis to the interacting
holographic dark energy model.

The expansion rate of the universe is described by the Hubble
parameter $H$, and the rate of acceleration/deceleration of the
expanding universe is characterized by the deceleration parameter
$q$. Furthermore, in order to find a more sensitive discriminator of
the expansion rate, let us consider the general expansion form for
the scale factor of the universe
\begin{equation}
a(t)=a(t_0)+\dot{a}|_0(t-t_0)+{\ddot{a}|_0\over
2}(t-t_0)^2+{{\stackrel{...}a}|_0\over
6}(t-t_0)^3+\dots.\label{aexp}
\end{equation}
Generically, various dark energy models give rise to families of
curves $a(t)$ having vastly different properties. In principle, we
can confine our attention to small value of $|t-t_0|$ in
(\ref{aexp}) because the acceleration of the universe is a fairly
recent phenomenon. Then, we see, following \cite{sahni}, that a new
diagnostic of dark energy dubbed statefinder can be constructed
using both second and third derivatives of the scale factor. The
second derivative is encoded in the deceleration parameter $q$, and
the third derivative is contained in the statefinder parameters
$\{r,s\}$. The statefinder parameters $\{r,s\}$ are defined as
 \begin{equation}
  \label{rs1}
 r\equiv\frac{\stackrel{...}a}{aH^3},~~~~~
 s\equiv\frac{r-1}{3(q-\frac{1}{2})}.
 \end{equation}
Note that the parameter $r$ is also called cosmic jerk. Thus the set
of quantities describing the geometry is extended to include $\{H,
q, r, s\}$. Trajectories in the $s-r$ plane corresponding to
different cosmological models exhibit qualitatively different
behaviors, so the statefinder can be used to discriminate different
cosmological models. The spatially flat LCDM (cosmological constant
$\lambda$ with cold dark matter) scenario corresponds to a fixed
point in the diagram
\begin{equation}
\{s,r\}\bigg\vert_{\rm LCDM} = \{ 0,1\} ~.\label{lcdm}
\end{equation}
Departure of a given dark energy model from this fixed point
provides a good way of establishing the ``distance'' of this model
from spatially flat LCDM \cite{sahni}. As demonstrated in Refs.
\cite{Alam:2003sc}--\cite{Chang:2007jr}, the statefinder can
successfully differentiate between a wide variety of dark energy
models including the cosmological constant, quintessence, phantom,
quintom, the Chaplygin gas, braneworld models and interacting dark
energy models, etc.. We can clearly identify the ``distance'' from a
given dark energy model to the LCDM scenario by using the $r(s)$
evolution diagram. The current location of the parameters $s$ and
$r$ in these diagrams can be calculated in models. The current
values of $s$ and $r$ are evidently valuable since we expect that
they can be extracted from data coming from SNAP (SuperNovae
Acceleration Probe) type experiments. Therefore, the statefinder
diagnosis combined with future SNAP observations may possibly be
used to discriminate between different dark energy
models.\footnote{It should be noted that the opinion of other
authors may not be so optimistic, see, e.g. \cite{Cattoen:2007id}}.
The statefinder parameter-pair also can be expressed as
\begin{equation}
 r=1+\frac{9(\rho+P)\dot{P}}{2\rho\dot{\rho}},~~~~~
 s=\frac{(\rho+P)\dot{P}}{P\dot{\rho}},
\end{equation}
where $\rho$ is the total density and $P$ is the total pressure.
Then, by using the Friedmann equation, we can obtain the following
concrete expressions
\begin{eqnarray}
 r&=&1-\frac{3}{2}\Omega_{\rm de}w'+
 3\Omega_{\rm de}w\left(1-\frac{1}{c}\sqrt{\Omega_{\rm de}}\right),\label{r}\\
s&=&1+w-\frac{w'}{3w}+\frac{b^2}{\Omega_{\rm de}}.\label{s}
\end{eqnarray}
Directly, from Eq. (\ref{w}), we have
\begin{eqnarray}
w'&=&\frac{\Omega_{\rm de}'}{\Omega_{\rm de}^2}
\left(b^2-\frac{1}{3c}\Omega_{\rm de}^{3/2}\right)\nonumber\\
&=&(1-\Omega_{\rm de})\left(b^2-\frac{\Omega_{\rm
de}^{3/2}}{3c}\right) \left[\frac{1}{\Omega_{\rm de}}-\frac{3b^2}
{\Omega_{\rm de}(1-\Omega_{\rm de})}+\frac{2}{c\sqrt{\Omega_{\rm
de}}}\right],\label{w1}
\end{eqnarray}
where the prime denotes the derivative with respect to $x=\ln a$.
Note that the whole dynamics of the universe in the interacting
holographic dark energy model is governed by the differential
equation (\ref{5}). So by solving Eq. (\ref{5}) we can get the
evolution solution of $\Omega_{\rm de}$ and then hold all the
cosmological quantities of interest and the whole dynamics of the
universe.

\begin{figure}[hbtp]
\begin{center}
\includegraphics[width=10cm]{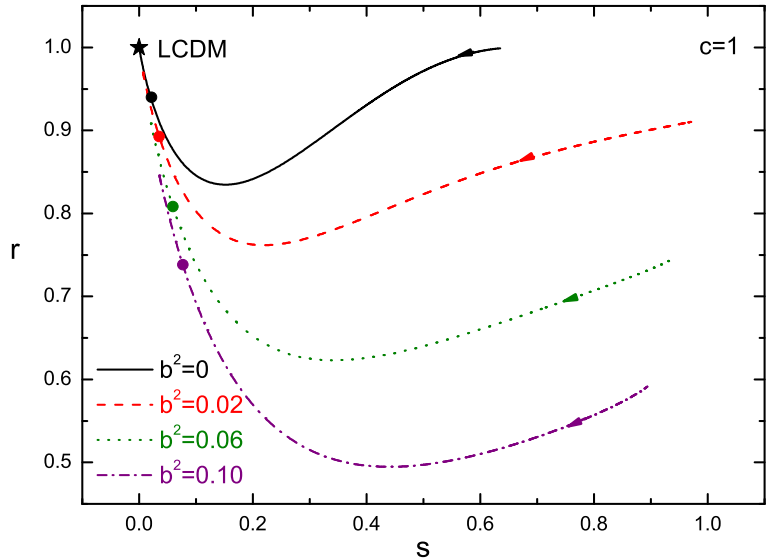}
\end{center}
\caption{The statefinder diagrams $r(s)$ for the interacting
holographic dark energy with a fixed parameter $c$ and different
coupling $b^2$. Selected curves of $r(s)$ are plotted by fixing
$c=1$, $\Omega_{\rm de0}=0.73$ and varying $b^2$ as 0, 0.02, 0.06
and 0.10, respectively. A star denotes the LCDM fixed point $(0,
1)$. The dots show today's values for the statefinder parameters
$(s_0, r_0)$.}
\end{figure}

\begin{figure}[htbp]
\centering $\begin{array}{c@{\hspace{0.2in}}c}
\multicolumn{1}{l}{\mbox{}} &
\multicolumn{1}{l}{\mbox{}} \\
\includegraphics[scale=0.8]{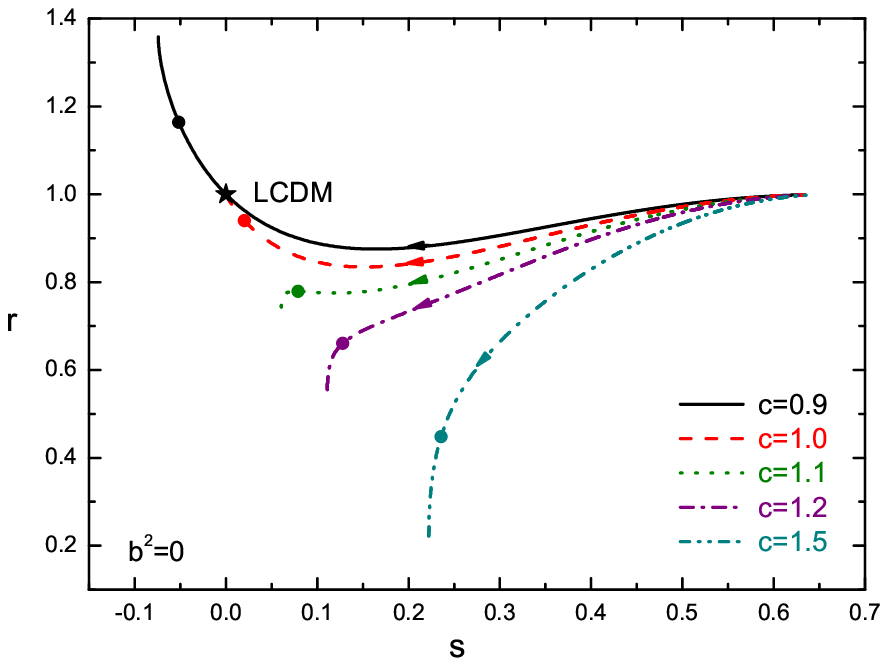} &\includegraphics[scale=0.8]{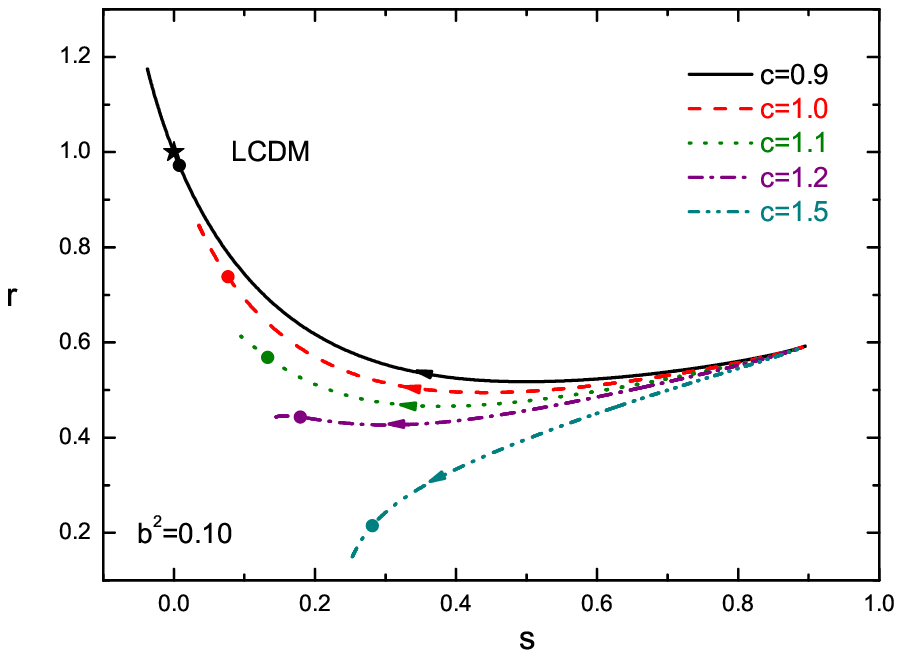} \\
%\mbox{\bf (b)} & \mbox{\bf (c)}
\end{array}$
\caption{The statefinder diagrams $r(s)$ for the holographic dark
energy model with different values of parameter $c$. In the plot, we
take $\Omega_{\rm de0}=0.73$ and vary $c$ as 0.9, 1.0, 1.1, 1.2 and
1.5, respectively. The left panel is for the holographic model
without interaction between the dark energy and dark matter
$(b^2=0)$, while the right one is for the case including the
interaction $(b^2=0.10)$. A star denotes the LCDM fixed point $(0,
1)$. The dots show today's values for the statefinder parameters
$(s_0, r_0)$.}
\end{figure}

In what follows we shall diagnose the interacting holographic dark
energy model employing the statefinder method. We shall analyze the
cases with fixed coupling constant $b^2$ and with fixed parameter
$c$, respectively. As demonstrated above, the information of this
model can be acquired by solving the differential equation
(\ref{5}). Making the redshift $z$ vary in a large enough range
involving far future and far past, %e.g. from $-1$ to the order of
%several hundreds,
one can solve the differential equation (\ref{5})
numerically and then get the evolution trajectories in the
statefinder $s-r$ planes for this model. For instance, we plot the
statefinder diagram in Fig. 3 for the cases of $c=1$ with various
values of coupling such as $b^2=0$, $0.02$, $0.06$ and $0.10$,
meanwhile the present density parameter of dark energy is taken to
be $\Omega_{\rm de0}=0.73$. The case $b^2=0$ corresponds to the
holographic dark energy model without interaction between dark
energy and dark matter. The arrows in the diagram denote the
evolution directions of the statefinder trajectories and the star
corresponds to $\{r=1,s=0\}$ representing the LCDM model. This
diagram shows that the evolution trajectories with different
interaction strengths exhibit different features in the statefinder
plane. When the interaction is absent, the $r(s)$ curve for
holographic dark energy ends at the LCDM fixed point, i.e., the
universe of this case will evolve to the de Sitter phase in the far
future. However, taking the interaction into account, the endpoints
of the $r(s)$ curves could not arrive at the LCDM fixed point $(0,
1)$, though all of the evolution trajectories tend to approach this
point. It should be mentioned that the statefinder diagnosis for
holographic dark energy model without interaction has been
investigated in detail in \cite{Zhang:2005holo}, where the focus is
put on the diagnosis of the different values of parameter $c$. The
statefinder analysis on the holographic dark energy in a non-flat
universe see \cite{Setare:2006xu}. In \cite{Zhang:2005holo}, it has
been demonstrated that from the statefinder viewpoint $c$ plays a
significant role in this model and it leads to the values of $\{r,
s\}$ in today and future tremendously different. In this paper, by
far, we have clearly seen that the interaction between holographic
dark energy and dark matter makes the statefinder evolutionary
trajectories with the same value of $c$ tremendously different also.
If the accurate information of $\{r_0, s_0\}$ can be extracted from
the future high-precision observational data in a model-independent
manner, these different features in this model can be discriminated
explicitly by experiments, one thus can use this method to test the
holographic dark energy model as well as other dark energy models.
Hence, today's values of $\{r, s\}$ play a significant role in the
statefinder diagnosis. We thus calculate the present values of the
statefinder parameters for different cases in the interacting
holographic dark energy model and mark them on evolution curves with
dots. It can be seen that stronger interaction results in longer
distance to the LCDM fixed point. The interaction between
holographic dark energy and dark matter prevents the holographic
dark energy from behaving as a cosmological constant $\lambda$
ultimately in the far future.

We also plotted the statefinder diagram in the $s-r$ plane for
different values of parameter $c$ with $b^2=0$ and $0.10$ in Fig. 4.
The left panel is for the holographic dark energy without
interaction while the right one is for the case involving the
interaction. The star in the figure also corresponds to the LCDM
fixed point and the dots marked on the curves represent the present
values of the statefinder parameters. Note that the true values of
$(s_0, r_0)$ of the universe should be determined in a
model-independent way, we can only pin our hope on the future
experiments to achieve this. We strongly expect that the future
high-precision experiments (e.g. SNAP) may provide sufficiently
large amount of precise data to release the information of
statefinders $\{H, q, r, s\}$ in a model-independent manner so as to
supply a way of discriminating different cosmological models with or
without dark energy. From Fig. 4, we can learn that the $r(s)$
evolutions have the similar behavior, i.e. the curves almost start
from a fixed point for both cases in the $s-r$ plane. Evidently, the
interaction between dark components makes the value of $r$ smaller
and the value of $s$ bigger. Also, obviously, the parameter $c$
plays a crucial role in the holographic model.

\begin{figure}[htbp]
\centering $\begin{array}{c@{\hspace{0.2in}}c}
\multicolumn{1}{l}{\mbox{}} &
\multicolumn{1}{l}{\mbox{}} \\
\includegraphics[scale=0.8]{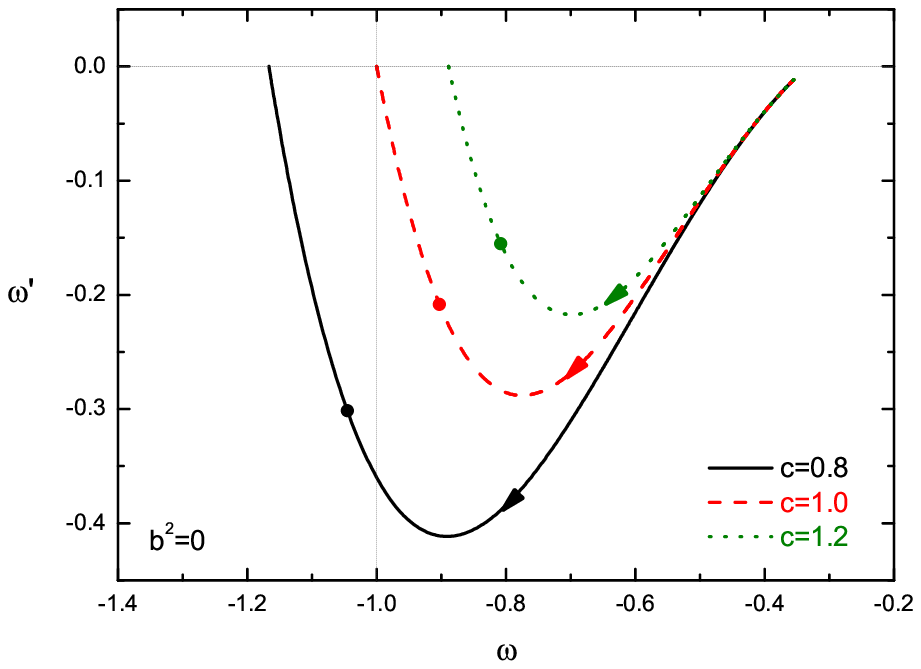} &\includegraphics[scale=0.8]{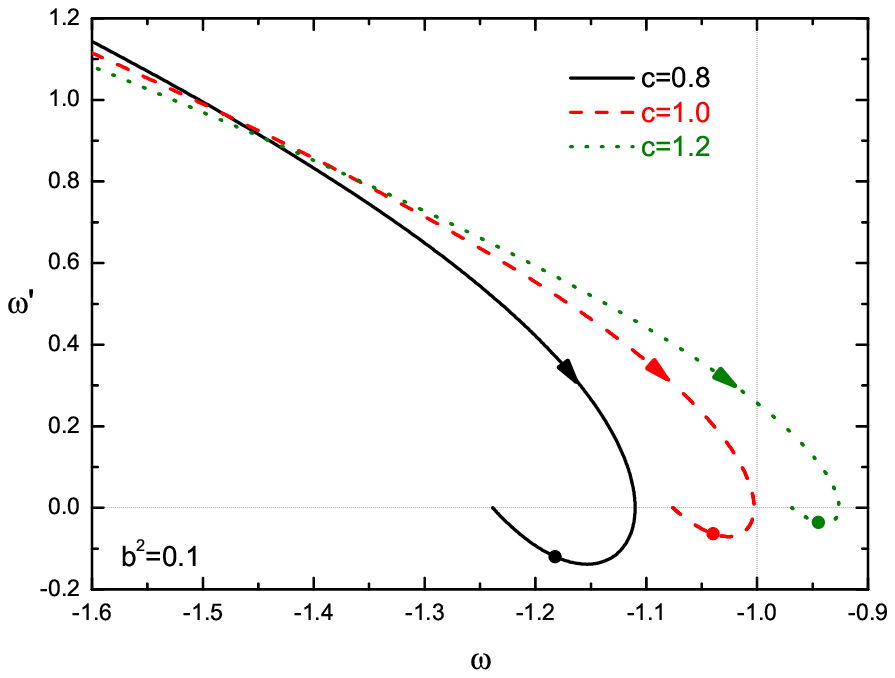}
%\mbox{\bf (b)} & \mbox{\bf (c)}
\end{array}$
\caption{The evolution trajectories of the holographic dark energy
for the cases with and without interaction in the $w-w'$ plane. The
coupling $b^2$ is taken to be 0 and 0.10, respectively. Selected
curves are plotted by taking $\Omega_{\rm de0}=0.73$ and varying $c$
as 0.8, 1.0 and 1.2, respectively. The dots denote the present
values of $(w, w')$.}
\end{figure}

As a complement to statefinder diagnosis, we investigate the
dynamical property of the interacting holographic dark energy in the
$w-w'$ phase plane, where $w'$ represents the derivative of $w$ with
respect to $\ln a$. Recently, this method became somewhat popular
for analyzing dark energy models. Caldwell and Linder
\cite{Caldwell:2005tm} proposed to explore the evolving behavior of
quintessence dark energy models and test the limits of quintessence
in the $w-w'$ plane, and they showed that the area occupied by
quintessence models in the phase plane can be divided into thawing
and freezing regions. Then, the method was used to analyze the
dynamical property of other dark energy models including more
general quintessence models \cite{Scherrer:2005je}, phantom models
\cite{Chiba:2005tj} and quintom models \cite{Guo:2006pc}, etc.. The
$w-w'$ analysis undoubtedly provides us with an alternative way of
classifying dark energy models using the quantities describing the
dynamical property of dark energy. But, it is obviously that the
$(w, w')$ pair is related to statefinder pair $(s, r)$ in a definite
way, see Eqs. (\ref{r}) and (\ref{s}). The merit of the statefinder
diagnosis method is that the statefinder parameters are constructed
from the scale factor $a$ and its derivatives, and they are expected
to be extracted in a model-independent way from observational data,
although it seems hard to achieve this at present. While the
advantage of the $w-w'$ analysis is that it is a direct dynamical
diagnosis for dark energy. Hence, the statefinder $s-r$ geometrical
diagnosis and the $w-w'$ dynamical diagnosis can be viewed as
complementarity in some sense.

Now let us investigate the interacting model of holographic dark
energy in the $w-w'$ plane. In Fig. 5, we plot the evolutionary
trajectories of the holographic dark energy in the $w-w'$ plane
where the selected curves correspond to $c=0.8$, $1.0$ and $1.2$,
respectively. The left graph is an illustrative example without
interaction to which we can compare the evolution of the interacting
holographic dark energy in the right diagram. Fig. 5 shows clearly
that the parameter $c$ and the interaction $b^2$ both play important
roles in the evolution history of the universe. The left graph tells
us: $c\geq1$ makes the holographic dark energy behave as
quintessence-type dark energy with $w\geq-1$ and $c<1$ makes the
holographic dark energy behave as quintom-type dark energy with $w$
crossing $-1$ during the evolution history. However, when the
interaction between dark components is present, the situation
becomes somewhat ambiguous because that the equation of state $w$
loses the ability of classifing dark energies definitely, due to the
fact that the interaction makes dark energy and dark matter be
entangled in each other. In this circumstance, the conceptions such
as quintessence, phantom and quintom are not so clear as usual. But,
anyway, we can still use these conceptions in an undemanding sense.
It should be noted that when we refer to these conceptions the only
thing of interest is the equation of state $w$. The right panel of
Fig. 5 tells us: with the interaction (a case of strong coupling,
$b^2=0.10$), $c\leq1$ makes the holographic dark energy behave as
phantom-type dark energy with $w\leq -1$ and $c>1$ makes the
holographic dark energy behave as quintom-type dark energy with $w$
crossing $-1$ during the evolution history.\footnote{It should be
noted that the old version of holographic cosmology is not
compatible with the phantom energy, see e.g.,
\cite{Bak:1999hd,Flanagan:1999jp}} In this diagram, the effect of
the interaction is shown again. When the coupling between the two
components is absent, the value of $w'$ first decreases from zero to
a minimum then increases again to zero meanwhile the value of $w$
decreases monotonically. Nevertheless, for the case involving the
interaction, $w$ increases first to a maximum and then decreases
meanwhile $w'$ decreases from a maximum to a negative minimum first
and then increases to zero again. Therefore, we see that the $w-w'$
dynamical diagnosis can provide us with a useful complement to the
statefinder geometrical diagnosis.

In summary, we have studied the interacting holographic dark energy
model from the statefinder viewpoint in this paper. Since the
accelerated expansion of the universe was found by astronomical
observations, many cosmological models involving dark energy
component or modifying gravity have been proposed to interpret this
cosmic acceleration. This leads to a problem of how to discriminate
between these various contenders. The statefinder diagnosis provides
a useful tool to break the possible degeneracy of different
cosmological models by constructing the parameters $\{r, s\}$ using
the higher derivative of the scale factor. Thus the method of
plotting the evolutionary trajectories of dark energy models in the
statefinder plane can be used to as a diagnostic tool to
discriminate between different models. Furthermore, the values of
$\{r, s\}$ of today, if can be extracted from precise observational
data in a model-independent way, can be viewed as a discriminator
for testing various cosmological models. On the other hand, though
we are lacking an underlying theory of the dark energy, this theory
is presumed to possess some features of a quantum gravity theory,
which can be explored speculatively by taking the holographic
principle of quantum gravity theory into account. So the holographic
dark energy model provides us with an attempt to explore the essence
of dark energy within a framework of fundamental theory. In
addition, some physicists believe that the involving of interaction
between dark energy and dark matter leads to some alleviation and
more understanding to the coincidence problem. It is thus worthwhile
to investigate the interacting model of holographic dark energy. We
analyzed the interacting holographic dark energy model employing the
statefinder parameters as a diagnostic tool. The statefinder
diagrams show that the interaction between dark sectors can
significantly affect the evolution of the universe and the
contributions of the interaction can be diagnosed out explicitly in
this method. At last, as the complement to the statefinder
geometrical diagnosis, a dynamical diagnosis was also studied, which
diagnoses the dynamical property of the interacting holographic dark
energy in the $w-w'$ phase plane. We hope that the future
high-precision observations can offer more and more accurate data to
determine these parameters precisely and consequently shed light on
the essence of dark energy.

\section*{Acknowledgements}
%\acknowledgments

This work was supported by the grants from the China Postdoctoral
Science Foundation (20060400104), the K. C. Wong Education
Foundation (Hong Kong), the National Natural Science Foundation of
China (10573003,10705041), and the National Basic Research Program
of China (2003CB716300).

%%%%%%%%%%%%%%%%%%%%%%%%%%%%%%%%%%%%%%%

\end{document}